\documentclass[11pt, oneside]{article}   	
\usepackage{geometry}                		
\usepackage{graphicx}				
								
\usepackage{hyperref}

\usepackage{amssymb}
\usepackage{amsmath}
\usepackage{color}
\usepackage{amsthm}
\usepackage{bbm}
\usepackage{soul,xcolor}
\setstcolor{red}

\newtheorem{remark}{ Remark}

\newtheorem{theorem}{ Theorem}[section]

\newtheorem{definition}{Definition}[section]

\newtheorem{assumption}{Assumption}[section]
\newcommand{\E}{\mathbb E}
\newcommand{\F}{\mathcal F}

\newcommand{\VIX}{\mbox{VIX}}

\title{Consistent Time-Homogeneous Modeling of SPX and VIX Derivatives\footnote{Data Sharing and Data Accessibility: Data sharing is not applicable to this article as no new data were created or analyzed in this study.}~\footnote{This work was partially supported by NSF grant DMS-1907518.}}

\author{A. Papanicolaou\thanks{Department of Mathematics, North Carolina State University, 2311 Stinson Drive, Raleigh, NC 27695. {\em apapani@ncsu.edu}}}

\begin{document}
\maketitle
\begin{abstract}
This paper shows how to recover a stochastic volatility model (SVM) from a market model of the VIX futures term structure. Market models have more flexibility for fitting of curves than do SVMs, and therefore are better suited for pricing VIX futures and VIX derivatives. But the VIX itself is a derivative of the S\&P500 (SPX) and it is common practice to price SPX derivatives using an SVM. Therefore, consistent modeling for both SPX and VIX should involve an SVM that can be obtained by inverting the market model. This paper's main result is a method for the recovery of a stochastic volatility function by solving an inverse problem where the input is the VIX function given by a market model. Analysis will show conditions necessary for there to be a unique solution to this inverse problem. The models are consistent if the recovered volatility function is non-negative. Examples are presented to illustrate the theory, to highlight the issue of negativity in solutions, and to show  the potential for inconsistency in non-Markov settings.
\textbf{AMS Subject Codes:} 91B24, 91B70, 45Q05
\end{abstract}
\tableofcontents


\section{Motivation and Formulation}
\label{sec:intro}
Volatility trading has increased in the 21st century with the introduction of derivatives on the VIX index. Two such derivatives are VIX futures, which began trading on the CBOE in 2004, and VIX (European) options, which began trading on the CBOE in 2006. There are also exchange traded notes (ETNs) written on these futures, and options written on these ETNs. Pricing of VIX derivatives uses so-called market models, which are stochastic models for the futures term structure. One thing to keep in mind when using a market model is that the published VIX index is computed from European S\&P500 (SPX) options, which means that the VIX is really an SPX derivative. Therefore, VIX prices may be partially determinable if there are established prices for SPX options. Moreover, pricing of SPX derivatives uses stochastic volatility models (SVMs) rather than market models. Hence, there is potential for conflicting prices if market models and SVMs are being used simultaneously to price VIX and SPX derivatives, respectively.

To understand why such a conflict would be a problem, consider a situation where a single financial institution has two separate trading floors: one for SPX derivatives and another for VIX derivatives. Each floor has its own traders who quote prices from their own respective models. If these models have substantially different assessments on the outcome of 30-day variance, then there could be inter-desk arbitrage, i.e., mispricings that allow for a third party (external to the institution) to take SPX derivative prices offered by the SPX desk and arbitrage them against VIX derivative prices offered by the VIX desk. A solution to this problem should provide a criterion for consistency of the models, and in a practical setting should provide a method for specification of one model in terms of the other. This paper presents such a solution.

\subsection{Problem Formulation}
Let $S_t$ denote the scalar price process for the SPX, and let $X_t$ denote a $d$-dimensional factor process with $d$ a positive integer. Consider a model where SPX returns are given by a risk-neutral SVM,
\begin{align}
\label{eq:SVM_dS}
\frac{dS_t}{S_t}&=rdt+v(X_t) dB_t\ ,\\
\label{eq:SVM_dX}
dX_t&=\mu(X_t)dt+\sigma(X_t)dW_t\ ,
\end{align}
where $r\geq 0$ is the risk-free rate, $v(x)$ is a scalar-valued volatility function, $\mu(x)$ is a $d$-dimensional drift,  $\sigma(x)$ is a $d\times d$ diffusion matrix, $W_t$ is a $d$-dimensional risk-neutral vector-valued Standard Brownian motion, and $B_t$ is a risk-neutral scalar Brownian motion, with correlations between $W_t$ and $B_t$ denoted with $\rho$,
\[dB_tdW_t^i= \rho_i dt\qquad\hbox{for }1\leq i\leq d\ .\]
Denote by $(\mathcal F_t)_{t\geq 0}$ the filtration generated by $W_t$ and $B_t$. The VIX is the square root of the risk-neutral expected realized 30-day variance, which in the continuous diffusion model of \eqref{eq:SVM_dS} and \eqref{eq:SVM_dX} is
\[\VIX_t = \sqrt{\E\left[ \frac{1}{\tau}\int_t^{t+\tau}v^2(X_u)du\Big|\F_t\right]}\ ,\]
where $\tau=$ 30 days, and with a VIX future being given by
\begin{equation}
\label{eq:vixfuture}
h_{T-t}(X_t) = \E[\VIX_T|\F_t]\qquad\hbox{for }t\leq T\ .
\end{equation}
For this SVM it is clear that the asset price $S_t$, the VIX, and all VIX futures are $\F_t$-adapted Markov processes. 

Separate from SVMs are market models that are designed to describe directly the VIX and VIX futures. Let $F_{t,T}$ be a market model's price for a VIX future with maturity $T$ at time $t\leq T$. These prices come from the following system of SDEs,
\begin{equation}
\label{eq:marketModel}
\frac{dF_{t,T}}{F_{t,T}} = \nu(t,T)dW_t\ ,
\end{equation}
where $W_t$ is the same Brownian motion from equation \eqref{eq:SVM_dX}, and where the volatility $\nu(t,T)$ is an $\mathcal F_t$-adapted $d$-dimensional row vector function that is specific for a given $T$. The model is applied simultaneously for multiple or a continuum of $T$'s, thereby forming an entire curve of VIX futures. It is important to keep in mind that equation \eqref{eq:marketModel} is generally a time inhomogeneous and non-Markovian model, but results in this paper apply to time-homogenous Markovian market models that are also driven by the factor process $X_t$ of equation \eqref{eq:SVM_dX}. Section \ref{sec:Consistency} will set forth assumptions for time homogeneity and Markovianity along with some explanation, but further discussion will come in Section \ref{sec:nonMarkov} where it will be shown how there is essentially a contradiction when trying to specify a consistent Markovian SVM with a non-Markovian market model. Market models considered in this paper include: a Bergomi-type market model\footnote{This paper throughout will refer to the ``Bergomi model" when perhaps it should say ``a Bergomi-type model for VIX futures", but this would be unwieldy so instead the name is used in a general sense. The original Bergomi model is for the curve of future instantaneous variances, not VIX directly.} where $\nu(t,T) = \gamma^* e^{-\mathbf k(T-t)} \sigma $ with $\mathbf k$ and $\sigma$ being $d\times d$ positive-definite matrices and $\gamma$ a $d \times 1$ volatility vector; a 3/2 market model where all futures are an expectation of a VIX given by $F_{t,t} = 1/X_t$ with $X_t$ being a Cox-Ingersol-Ross (CIR) process; a double Nelson (or double mean reverting model) similar to the model in \cite{gatheral2013}; a non-stationary model where $X_t$ is Brownian motion.

In practice, it is a good idea to use market models because VIX futures are very liquid with a richness of information for understanding the state of volatility. Therefore, it is sensible to first define a market model, and second to build an SVM with the structure of the market model taken into consideration. If the market model is Markovian with the same factors as the SVM, then the instantaneous variance $v^2(x)$ is the solution to an inverse problem. The formulation of this inverse problem is as follows: if the coefficients of the factor process $\mu(x)$ and $\sigma(x)$ are known for all $x$, and the market model has provided the function $h_0(x)$, then the inverse problem for $v^2(x)$ is expressed as
\begin{equation}
\label{eq:inverseProblemIntro}
h_0^2(x) = \E\left[ \frac{1}{\tau}\int_t^{t+\tau}v^2(X_u)du\Big|X_t= x\right] \ .
\end{equation} 
In Section \ref{sec:consisency} it is shown that if the process $X_t$ is ergodic and has an invariant law relative to which its infinitesimal generator is symmetric, making $X_t$ reversible, and has also a spectral gap \cite{bakry2013analysis}, then under some integrability conditions for $h_0^2$ there exists a unique $v^2(x)$ that is the solution to equation \eqref{eq:inverseProblemIntro}. If in addition the solution is non-negative, then the market model has provided a valid volatility function for equation \eqref{eq:SVM_dS}, leaving the correlation coefficients $\rho_1,\ldots,\rho_d$ as the only remaining parameters to be determined for the SVM. However, non-negativity of the solution is difficult to prove; this issue is explored further in the examples of Section \ref{sec:examples}. Similar non-negativity issues arise elsewhere in the literature, for example in quantum inverse scattering theory \cite{chadan2012inverse,chadan2007positivity} and in Fourier analysis \cite{tuck2006positivity}. For the most part, results in the literature proving positivity are less general than theory for existence of solutions.

\subsection{Background Literature}
Background for SVMs, including the Heston model, can be found in various books and papers, including \cite{Gatheral06,heston93}, and more general models involving jumps in \cite{duffie2000transform}. The Bergomi model for future variance is introduced in \cite{bergomi2005smile,bergomi2008smile,bergomi2015stochastic}, and a consistency condition for the drift in futures curves for variance is given in \cite{buehler2006consistent}. Term structure and the associated Heath-Jarrow-Morton (HJM) framework are discussed in \cite{carmona,HJM1992,RS1995,roncoroni}. In \cite{jacquier2018vix} a market model with stochastic volatility is used to derive formulae for VIX futures.   

In the past decade there has been a lot of research on joint models for SPX and VIX options, which includes some re-evaluation of widely-used SVMs and a search for new models to fit both markets. One such example is the so-called 3/2 model, which is analyzed in \cite{BB,Drimus11} and is a popular choice because it is able to reproduce the increasing right-hand implied-volatility skew in VIX options. The search for a model to simultaneously calibrate an SVM to both SPX and VIX options is done in \cite{gatheral2013,gatheral2008consistent} using a two-factor diffusion model; in \cite{cont2013consistent} a numerically efficient model for joint calibration is proposed using affine jump diffusions; in \cite{lin2010consistent} there is exploration of the Heston model with jumps revealing evidence in VIX options that suggests there are jumps in the volatility process; in \cite{kokholm2015joint} there is further exploration of the Heston model with jumps and the role played by the Feller condition; in \cite{pacati2018smiling} a displacement or volatility push-up is proposed to improve the joint fit of affine models; in \cite{papanicolaouSircar2014} a model for joint calibration is proposed using a regime-switching extension of the Heston model \cite{heston93}; in \cite{fouque2018heston} it is proposed to use a Heston vol-of-vol model. A different approach is taken in \cite{guyon2020joint}  wherein a joint distribution from a non-parametric family is fit satisfying both the marginal distributions from VIX and SPX options, and it is shown to be arbitrage free. Non-model-specific analysis of the joint SPX and VIX markets includes \cite{papanicolaou2018} and the data analysis of future variance-swap rates in \cite{papanicolaou2016}. The problem of consistency between SVMs and market models is also studied in the PhD thesis of Alex Badran \cite{BadranThesis}.

\subsection{Results and Organization of this Paper}

Section \ref{sec:Consistency} introduces Definition \ref{def:consistency}, stating formally the meaning of \textit{consistency} between an SVM and a market model. Section \ref{sec:inverseProblem} has the main result of this paper, which is a theorem for the solvability of equation \eqref{eq:inverseProblemIntro}.  Section \ref{sec:examples} has examples of tractable models: Sections \ref{sec:1factorBergomiPart1} and \ref{sec:multiFactorBergomi} explore the scalar and multivariate Bergomi models, respectively, for which the inverse problem has an explicit eigenfunction expansion; Section \ref{sec:3/2model} looks at the market model where $\VIX_t^2$ is a 3/2 process, which also has an explicit eigenfunction expansion; Section \ref{sec:doubleNelson} looks at the double Nelson market model that is tractable with nice statistical features to fit the data but with an inverse problem that does not have a positive solution for all $x$; Section \ref{sec:brownianMotion} looks at a non-stationary model with a Brownian motion factor process for which an application of Bochner's theorem ensures non-negativity. Section \ref{sec:nonMarkov} has further discussion on the issue of non-Markovian market models.

\section{Definitions and Main Result}
\label{sec:consisency}
The value of a future contract with a fixed horizon is referred to as a \textit{constant maturity future} (CMF), that is, for a constant $\theta\geq 0$ the CMF price with horizon $\theta$ given by the SVM is,
\begin{equation}
\label{eq:cmfvix}
h_\theta(X_t)= \E[\VIX_{t+\theta}|\F_t]\ ,
\end{equation}
and the CMF given by the market model is,
\begin{equation}
\label{eq:marketfuture}
F_{t,t+\theta} = \E[F_{t+\theta,t+\theta}|\F_t]\ .
\end{equation}
Unlike regular futures, CMFs are not risk-neutral martingales. Instead, their differential has a non-zero drift,
\begin{equation}
\label{eq:CMF_dF}
\frac{dF_{t,t+\theta}}{F_{t,t+\theta}} = Y_t^\theta dt+\nu(t,t+\theta)dW_t\ ,
\end{equation}
where $Y_t^\theta=\frac{\partial}{\partial T}\log(F_{t,T})\Big|_{T=t+\theta}$. In the stationary case there is the following representation of $Y_t^\theta$,
\begin{align}
\label{eq:Ydef}
Y_t^\theta &=\left[-\int_{-\infty}^t\nu(u,T)\frac{\partial}{\partial T}\nu^*(u,T)du+\int_{-\infty}^t\frac{\partial}{\partial T}\nu(u,T)dW_u\right]_{T=t+\theta}\ ,
\end{align}
from which it is seen that stationarity requires some integrability of $\nu(t,t+\theta)$ and $\frac{\partial}{\partial T}\nu(t,T)\Big|_{T=t+\theta}$. For example, $\nu(t,T) = \gamma e^{-k(T-t)}$ leads to stationary CMFs. The quantity $Y_t^\theta$ has a financial significance because it is the roll-yield of a trading strategy to track the CMF's returns (see \cite{avellanedaPapanicolaou2017}).

\subsection{Consistency}
\label{sec:Consistency}

It is first necessary to define this paper's meaning for consistency:
\begin{definition}[Consistency] 
\label{def:consistency}
 Assume there exists a unique strong solution for the SDE appearing in equation \eqref{eq:SVM_dX} of the SVM, and there also exists unique strong solutions for the SDEs for CMFS given by \eqref{eq:CMF_dF}. The SVM and the market model have consistent prices if the CMFs agree, that is, if
\begin{equation}
\label{eq:noArbEquation}
F_{t,t+\theta} = h_\theta(X_t) \qquad\hbox{a.s. $\forall t\geq 0$ and $\forall \theta\geq 0$}\ ,
\end{equation}
where $h_\theta(x)$ is the SVM's CMF as defined in \eqref{eq:cmfvix}.
\end{definition}

\begin{remark}
The essential step in confirming consistency between an SVM and a market model is to prove the statement expressed by equation \eqref{eq:noArbEquation}. However, the SVM is such that $h_{T-t}(X_t) = \E[h_0(X_T)|\mathcal F_t]$ for all $t\geq 0$ and the market model is such that $F_{t,T} = \E[F_{T,T}|\F_t]$ for all $T\geq t$, and so it is sufficient to show 
\[F_{t,t} = h_0(X_t)\qquad\hbox{a.s. $\forall t\geq0$}\ .\]
That is, the SVM and the market model share the same filtration $(\mathcal F_t)_{t\geq 0}$, and both $h_{T-t}(X_t)$ and $F_{t,T}$ are martingales by construction, and so all that needs to be checked is that the models have agreement between their respective VIX processes.
\end{remark}

If Definition \ref{def:consistency} does not hold then there is potential for arbitrage. For example in \cite{papanicolaou2018} it is shown how prices should respect certain structural bounds, otherwise there are options portfolios that produce arbitrage. From Definition \ref{def:consistency}, the first thing to notice is that solutions to equation \eqref{eq:CMF_dF} are functions of $X_t$ and do not depend separately on $t$. Thus, Markovian SVM future prices need to be equal to those of the market model, and it stands to reason that the market model should also be a Markov process. Therefore, the simplest approach is to assume that both the SVM and the market model are Markovian and both driven by the factor process $X_t$; discussion and a counterexample related to this issue will come in Section \ref{sec:nonMarkov}.

The second thing to notice from Definition \ref{def:consistency} is that time homogeneity in the SVM implies time homogeneity in the market model. The reason being, that the differential of $h_\theta(X_t)$ obtained from It\^o's lemma and the SDE in \eqref{eq:SVM_dX} have time-independent coefficients, and therefore the differential of $F_{t,t+\theta}$ must also have time-independent coefficients. 

\begin{assumption}[Time-Homogeneous Markovian Market Model Driven by $X_t$]
\label{assumption:markovX}
The market model with CMFs given by equation \eqref{eq:CMF_dF} is a Time-Homogeneous Markov model driven by the same factor process as the SVM. In particular, the CMF $F_{t,t+\theta}$ has roll-yield functions $f_\theta(x)$ and volatility row vector functions $\nu_\theta(x)$ for each horizon $\theta\geq0$, such that the CMF dynamics are,
\begin{equation}
\label{eq:CMF_TH}
\frac{dF_{t,t+\theta}}{F_{t,t+\theta}} = f_\theta(X_t) dt+\nu_\theta(X_t)dW_t\ ,
\end{equation}
where $X_t$ is the factor process given by equation \eqref{eq:SVM_dX}. Moreover, there is an initial curve $F_{0,\theta}$ such that $F_{0,\theta}=\E[F_{\theta,\theta}|\mathcal F_0]$ for all $\theta\geq 0$, and there is an initial value $X_0$ such that $X_0\in h_0^{-1}(F_{0,0})$ a.s.
\end{assumption}
At this point it is appropriate to formally state sufficient conditions on the SDE coefficients.
\begin{assumption}
    \label{assumption:strongSolutions} Coefficients $\mu(x)$ and $\sigma(x)$ are globally Lipschitz continuous so that equation \eqref{eq:SVM_dX} has a strong solution. Coefficients $f_\theta(x)$ and $\nu_\theta(x)$ are bounded so that \eqref{eq:CMF_TH} and \eqref{eq:dF_TH} have unique strong solutions and the futures of \eqref{eq:dF_TH} are true martingales.
\end{assumption}

From Assumptions \ref{assumption:markovX} and \ref{assumption:strongSolutions} it is assured that Definition \ref{def:consistency} is meaningful. Assumption \ref{assumption:markovX} is necessary for the theory in this paper, but Assumption \ref{assumption:strongSolutions} is not always needed. Indeed, there are several important non-Lipschitz examples, such as the Heston SVM, or various other models where $X_t$ is a CIR process (see the 3/2 model of Section \ref{sec:3/2model}). Assumption \ref{assumption:strongSolutions} asserts boundedness of $f_\theta(x)$ and $\nu_\theta(x)$, but a less restrictive criterion is for $f_\theta(x)$ to allow a well-defined Riemann integral $\int_0^tf_\theta(X_u)du$ and for $\nu_\theta(x)$ to satisfy the Novikov condition, 
\[\E\exp\left(\frac{1}{2}\int_0^T\|\nu_{T-t}(X_t)\|^2dt\right)<\infty\qquad\hbox{for all }0\leq T<\infty\ .\]

Assumption \ref{assumption:markovX} says that the roll-yields from equation \eqref{eq:CMF_dF} are now functions of $X_t$, namely, $Y_t^\theta=f_\theta(X_t)$ for all $\theta\geq 0$. Given Assumption \ref{assumption:markovX}, the market model's future dynamics can be rewritten as
\begin{equation}
    \label{eq:dF_TH}
    \frac{dF_{t,T}}{F_{t,T}}=\frac{dF_{t,t+\theta}}{F_{t,t+\theta}}\Bigg|_{\theta=T-t}-f_{T-t}(X_t)dt=\nu_{T-t}(X_t)dW_t\ ,
\end{equation}
Under Assumptions \ref{assumption:markovX} and \ref{assumption:strongSolutions}, It\^o's lemma can be applied to check whether or not a model satisfies the consistency of Definition \ref{def:consistency} and equation \eqref{eq:noArbEquation}. Indeed, denote the operator $\mathcal L$,
\begin{equation}
\label{eq:infgen}
\mathcal L = \frac{1}{2}\hbox{trace}[\sigma \sigma^*(x)\nabla\nabla^*]  +\mu^*(x)\nabla\ ,
\end{equation}
which is the infinitesimal generator of the factor process $X_t$. If $h_\theta(x)$ has sufficient differentiability, then $dh_\theta(X_t)$ is set equal to the right-hand side of \eqref{eq:CMF_TH} to obtain the following pair of consistency equations, 
\begin{eqnarray}
\label{eq:noArb_driftMulti}
\mathcal{L} h_\theta(X_t)&=&f_\theta(X_t) h_\theta(X_t)\\
\label{eq:noArb_diffMulti}
\sigma^*(X_t)\nabla h_\theta(X_t)&=&\nu_\theta^*(X_t) h_\theta(X_t)\ ,
\end{eqnarray}
with the initial condition satisfying and $X_0\in h_0^{-1}(F_{0,0})$.
\begin{remark}[Buehler's Condition]
\label{remark:buehlerCondition}
Equation \eqref{eq:noArb_driftMulti} is Buehler's condition, which was identified for expected variance in \cite{buehler2006consistent}.
\end{remark}
\begin{remark}[Initializing Curve Models with Market Data]
    Practical use of market models often involves the insertion of VIX curve data as an initial condition. The advantage to this approach is that the model is able to directly assimilate futures prices observed in the market. If the initial curve $F_{0,\theta}$ is given, then the solution to equation \eqref{eq:CMF_TH} is 
    \[F_{t,t+\theta}=F_{0,\theta}\exp\left(\int_0^t\left(f_\theta(X_u)-\frac{1}{2}\|\nu_\theta(X_u)\|^2\right)du + \int_0^t\nu_\theta(X_u)dW_u\right)\ ,\]
    which may not be a time-homogeneous market model driven by $X_t$ if the initial conditions stated in Assumption \ref{assumption:markovX} are not satisfied, that is, time inhomogeneity could arise if the initial curve cannot be written as a function of $X_0$. Within the framework of Assumption \ref{assumption:markovX}, a time-homogeneous market model can be initialized with curve data $F_{0,\theta}$ if  $F_{0,\theta}=\E[F_{\theta,\theta}|\mathcal F_0]$ and if there exists $X_0$ with $X_0\in h_0^{-1}(F_{0,0})$. In practice, the dimension of $X_t$ should be sufficiently high so that $h_0^{-1}(F_{0,0})$ contains $X_0$, i.e., so that there exists $x$ in the domain of $X_0$ with $x\in h_0^{-1}(F_{0,0})$.
\end{remark}

\subsection{Main Result: Markovian Inverse Problem for $v^2(x)$}
\label{sec:inverseProblem}

Let $h(x)=h_0(x)$  denote the VIX. Suppose that $h(x)$ and the market model satisfy Definition \ref{def:consistency} and Assumption \ref{assumption:markovX}. A function $v^2(x)$ should be found for consistent specification of the SVM. If $h(x)$ is already known, then finding $v^2(x)$ amounts to solving an inverse problem,
\begin{equation}
\label{eq:inverseProblem}
h^2(x)= \E\left[ \frac{1}{\tau}\int_t^{t+\tau}v^2(X_u)du\Big|X_t=x\right]\ ,
\end{equation}
where a solution is a function $v^2:\mathbb R^d\rightarrow\mathbb R$ that satisfies equation \eqref{eq:inverseProblem}. This solution admits a valid SVM if it is non-negative.


\subsubsection{General Solvability}
\label{sec:generalSolvability}
The inverse problem can be solved for a general class of factor processes. Let the factor process $X_t$ be a stationary ergodic process with infinitesimal generator $\mathcal L$ given by \eqref{eq:infgen}. Let $\omega$ denote $X_t$'s invariant measure. Here and in the sequel, expectation with respect to the invariant measure for any (integrable) test function $g$ is denoted by
\[\left<g\right>:= \int g(x)d\omega(x)\ ,\]
and all calculations to come will follow the analytical framework of semigroups for diffusions defined in \cite{bakry2013analysis}. It will be necessary to assume existence of a unique invariant measure and that the operator $\mathcal L$ has a spectral gap:

\begin{assumption}[Unique Invariant Measure]
    \label{assumption:invariantDensity}
    There is a unique invariant measure $\omega$ such that $\left<\mathcal{L} g\right> =0$ for any test function $g(x)$.\footnote{Conditions for existence of a unique invariant \textit{measure} are given in \cite{pardouxVeretnnikov2001}. They include boundedness and uniform ellipticity of matrices $\sigma\sigma^*(x)$, and also that $\lim\sup_{\|x\|\rightarrow\infty}x^*\mu(x)\leq -c \|x\|^{1+\alpha}$ for some $c>0$ and $\alpha\geq -1$.} 
\end{assumption}

\begin{assumption}[Spectral Gap]
\label{assumption:spectralGap}
The operator $\mathcal L$ is symmetric, that is, $\left< g_1\mathcal{L}g_2\right>= \left< g_2\mathcal{L}g_1\right>$ for any test functions $g_1(x)$ and $g_2(x)$, with a spectrum that is non-positive with a gap at zero. In other words, there is a constant $\lambda>0$ such that,
\begin{equation}
\label{eq:spectralGap}
\left<(e^{\mathcal Lt}g)^2\right>\leq e^{-\lambda t}\left<g^2\right>\ ,
\end{equation}
for all $t\geq 0$ and for any $g(x)$ such that $\left<g\right> = 0$ and $\left<g^2\right><\infty$.
Here $e^{\mathcal Lt}g$ denotes the contraction semigroup generated by $\mathcal{L}$, and given by 
$$e^{\mathcal Lt}g(x)= \E\left[g(X_t)\Big|X_0=x\right]$$ for bounded $g(x)$ as well as for square integrable ones.
\end{assumption}
Clearly $|e^{\mathcal Lt}g(x)| \leq \sup_y |g(y)|$ and also $\left<(e^{\mathcal Lt}g)^2\right>\leq \left<g^2\right>$ for all suitable $g(x)$, $t\geq 0$.
Conditions on the symmetric diffusion generator $\mathcal{L}$ to have a spectral gap are given in \cite{bakry,bakry2013analysis}, with the Ornstein-Uhlenbeck generator
being the canonical case that motivates the more general theory\footnote{The theory of Pardoux and Veretennikov \cite{pardouxVeretnnikov2001} can also be used for Theorem \ref{thm:main}.}. The examples of Section \ref{sec:examples} explore further the scope of the theory.

\begin{theorem}[General Solvability of Inverse Problem]
\label{thm:main}
Assume $h^2(x)$ is such that $\left< h^4\right><\infty$ and $\left<(\mathcal Lh^2)^2\right><\infty$, where $\mathcal L$ is the operator from equation \eqref{eq:infgen}. Given Assumptions \ref{assumption:invariantDensity} and \ref{assumption:spectralGap}, a square-integrable solution to equation \eqref{eq:inverseProblem} exists.
\end{theorem}
\begin{proof}[Proof of Theorem \ref{thm:main}]
By writing the solution as $v^2(x) = \left<h^2\right>+\xi(x)$, the inverse problem of equation \eqref{eq:inverseProblem} can be rewritten as
\[h^2(x) -\left<h^2\right>= \Phi\xi(x) \ ,\]
where the operator $\Phi$ is defined by
\begin{equation}
\label{eq:Phi}
\Phi = \frac{1}{\tau}\int_0^\tau e^{\mathcal Lu}du.
\end{equation}
Using the invariant measure it is clear the solution $\xi$ is now centered,
\[\left<\xi\right> = 0\ ,\]
because $0=\left<h^2 -\left<h^2\right>\right>=\left<\Phi \xi\right> = \int (\Phi \xi) d\omega = \int \xi d\omega=\left<\xi\right>$, and the inverse problem is posed as
\begin{equation}
\label{eq:inverseProblemWithPhi}
\Phi \xi(x) = h^2(x)- \left<h^2\right>\ .
\end{equation}
The operator $\Phi$ is an averaging operator, and so it stands to reason that $h^2(x)$ is more regular than $\xi(x)$.  The operator $\mathcal L$ is applied to both sides of equation \eqref{eq:inverseProblemWithPhi}, and because by assumption the quantity $\mathcal{L}h^2(x)$ is well defined, it follows that
\[\mathcal L \Phi \xi(x) = \mathcal L h^2(x)\ .\] 
Using the algebraic properties of the semigroup operator (see \cite{bakry2013analysis,rudin}),
\[\mathcal L\Phi = \frac{1}{\tau}\int_0^\tau \mathcal Le^{\mathcal Lu}du = \frac{1}{\tau}\left(e^{\mathcal L\tau}-I\right)\ ,\]
which can be rearranged to obtain,
\begin{equation}
\label{eq:uninvertedSolution}
-\tau\mathcal Lh^2(x) = -\tau\mathcal L\Phi\xi(x) = (I-(I+\tau\mathcal L\Phi))\xi=(I-e^{\mathcal L\tau})\xi(x)\ ,
\end{equation}
and due to the spectral gap of Assumption \ref{assumption:spectralGap} the solution can be written with a (convergent) geometric series,
\[\xi(x) =-\tau\sum_{n=0}^\infty e^{n\mathcal L\mathcal \tau}\mathcal Lh^2(x)\ .\]
Note the solvability assumption: given the spectral gap there is a solution if and only if $\left<\mathcal L h^2\right>=0$, which is the same as equation \eqref{eq:solvability}
after applying consistency equations \eqref{eq:noArb_driftMulti} and \eqref{eq:noArb_diffMulti}. In addition, it is needed to use the fact that,
$$\mathcal{L}h^2(x) = 2h(x) \mathcal{L} h(x)+ \|\sigma^*(x) \nabla h(x) \|^2\ .$$
Uniqueness of a square integrable solution with $\left<\xi\right>=0$ also follows from equation \eqref{eq:uninvertedSolution}: for any two solutions $\xi(x)$ and $\xi'(x)$ having $\left<\xi^2\right>+\left<\xi'^2\right><\infty$ it must be that $\mathcal L\Phi \xi(x)=\mathcal L\Phi \xi'(x)$, or $(I-e^{\mathcal L\tau})\xi(x)=(I-e^{\mathcal L\tau})\xi'(x)$. By inverting the operator $I-e^{\mathcal L\tau}$ it is clear that $\xi(x)=\xi'(x)$ for a.e. $x$.

Multiplying both sides of equation \eqref{eq:uninvertedSolution} by $\xi(x)$ and taking brackets yields,
\[\left<\xi^2\right>=-\tau\left<\xi\mathcal Lh^2\right>+\left<\xi e^{\mathcal L\tau}\xi\right>\ .\]
From symmetry of $\mathcal L$ and the spectral gap in equation \eqref{eq:spectralGap} there is the following estimate,
\[\left<\xi e^{\mathcal L\tau}\xi\right> = \left<(e^{\mathcal L\tau/2}\xi)^2\right>\leq e^{-\lambda\tau/2}\left<\xi^2\right>\ ,\]
which is inserted into the previous equation to obtain,
\[\left<\xi^2\right>\leq -\tau\left<\xi\mathcal Lh^2\right>+e^{-\lambda\tau/2}\left<\xi^2\right>\ .\]
Rearranging and applying Cauchy-Schwartz yields the estimate,
\[(1-e^{-\lambda\tau/2})\left<\xi^2\right>\leq -\tau\left<\xi\mathcal Lh^2\right>\leq \tau \sqrt{\left<\xi^2\right>\left<(\mathcal Lh^2)^2\right>}\ , \]
which for $\lambda>0$ is rearranged to obtain an estimate on the norm of the solution,
\begin{equation}
\label{eq:xiEstimate}
\left<\xi^2\right>\leq \left(\frac{\tau}{1-e^{-\lambda\tau/2}}\right)^2\left<(\mathcal Lh^2)^2\right>\ .
\end{equation}
The bound \eqref{eq:xiEstimate} shows that the solution is square integrable against the invariant density, given our assumptions about $h^2(x)$
and the spectral gap. 
\end{proof}

\begin{remark}
The implication of Theorem \ref{thm:main} is that, if $h(x)$ is given by a market model and a unique solution to equation \eqref{eq:inverseProblem} is non-negative for a.e. $x$, and if equation \eqref{eq:SVM_dX} has a strong solution, then there is an SVM that is consistent in the sense of Definition \ref{def:consistency}.
\end{remark}

\begin{remark}
It may be the case that equation \eqref{eq:inverseProblem} is solvable but does not have a solution that is non-negative for a.e. $x$, even though it is denoted by $v^2(x)$ because that is how the problem is posed. In this case, for the proposed market model there does not exists an SVM that is consistent in the sense of Definition \ref{def:consistency}.
\end{remark}

\begin{remark}
If a square-integrable solution $v^2(x)$ to equation \eqref{eq:inverseProblem} exists, and if the SVM and market model are consistent (in the sense of Definition \ref{def:consistency}), then from the consistency equations of \eqref{eq:noArb_driftMulti} and \eqref{eq:noArb_diffMulti} there is the following solvability condition for the inverse problem,
\begin{equation}
\label{eq:solvability}
\left< (2f+  \| \nu \|^2)h^2\right>=0\ ,
\end{equation}
where $f(x)$ is the roll yield and $\nu(x)$ the volatility in \eqref{eq:CMF_TH} with $\theta=0$.
\end{remark}

\begin{remark}
The solvability condition in equation \eqref{eq:solvability} is analogous to the Fredholm alternative in finite Euclidean space (see \cite{bakry2013analysis,rudin}).
It is an integral condition that involves the roll yield $f(x)$ and volatility of the market model $\nu(x)$, the VIX $h(x)$, and the invariant measure of the factor process $\omega$.
\end{remark}

\begin{remark}[Symmetric Operators]
For $X_t$ given by equation \eqref{eq:SVM_dX}, if there is an invariant density $\omega(x)$, then the operator $\mathcal L$ of equation \eqref{eq:infgen} is symmetric if there are matrices $A(x)$ such that $\mathcal L$ can be written in self-adjoint form,
\[\mathcal L g(x)= \frac{1}{2\omega(x)}\nabla \cdot (A(x)\nabla g(x))\ ,\]
for any test function $g(x)$. In other words, $\sigma(x)$ and $\mu(x)$ need to satisfy
\begin{align*}
    A(x) &= \sigma\sigma^*(x)\omega(x)\\
    \nabla \cdot A(x)&=2\mu^*(x)\omega(x)\ .
\end{align*}
This shows us that the symmetry of Assumption \ref{assumption:spectralGap}
is somewhat restrictive. However, symmetry is not always required to solve the inverse problem, as will be shown in the examples of Section \ref{sec:examples}.
\end{remark}


\subsubsection{Solution via Eigenseries Expansion}
\label{sec:eigenMethod}
If the operator $\mathcal L$ has a complete basis of orthogonal eigenfunctions, then so does $\Phi$ given in \eqref{eq:Phi}, and then the solution to the inverse problem \eqref{eq:inverseProblem} can be found by computing eigencoefficients in a series expansion of $v^2(x)$. For many such cases there are transition densities for the factor process $X_t$ given $X_0$, and so equation \eqref{eq:inverseProblem} can be written using a kernel,
\[h^2(x) = \int \Phi(y,x)v^2(y)dy\ ,\]
where the kernel is,
\[\Phi(y,x) = \frac{1}{\tau}\int_0^\tau\frac{\partial}{\partial y}\mathbb P(X_u\leq y|X_0=x)du\ .\]
Suppose there are eigenfunctions $\psi_n:\mathbb R^d\rightarrow\mathbb R$ such that for an index value $n\in\{0,1,2,\dots\}$,
\[\int \Phi(y,x)\psi_n(y)dy = \lambda_n\psi_n(x)\ ,\]
where $\lambda_n\neq 0$, and suppose there is an invariant density $\omega(x)>0$ such that  any pair is orthogonal,
\[\int \psi_n(x)\psi_m(x)\omega(x)dx = \left<\psi_n^2\right> \delta(n-m)\ .\]
Suppose additionally that these eigenfunctions form a complete basis in $L^2(\mathbb R^d;\omega)$, i.e,. if $\left< v^4\right><\infty$ then there are coefficients $a_0,a_1 ,a_2 \dots$ such that,
\[v^2(x) = \sum_{n=0}^\infty  a_n \psi_n(x)\ ,~~~~~ \sum_{n=0}^\infty  |a_n|^2 < \infty \ .\]
If $v^2$ is the solution to the inverse problem then there is eigenseries expansion,
\[h^2(x) = \sum_{n=0}^\infty a_n  \lambda_n\psi_n(x), ~  \sum_{n=0}^\infty  |a_n \lambda_n|^2 < \infty \ , \]
and via orthogonality the $a_n$'s are solved for,
\[a_n = \frac{1}{ \lambda_n\left<\psi_n^2\right>}\int h^2(x)\psi_n(x)\omega(x)dx\ .\]
This provides a (unique) solution to equation \eqref{eq:inverseProblem}. 
\begin{remark}
    The eigenfunction expansion presented in this section can be reformulated without the assumption of a transition density; see \cite{linetsky2007spectral} for spectral theory of general semi-group operators.
\end{remark}


\section{Application to Tractable Models}
\label{sec:examples}
This section presents some examples of models that are applicable in practice, i.e., simulation, numerics, data calibration, etc., can be done within a reasonable amount of time. All models considered are Markov with strong SDE solutions, as per Assumptions \ref{assumption:markovX} and \ref{assumption:strongSolutions}. It will be assumed that $h_\theta(x)=F_{t,t+\theta}$ for all $\theta\geq 0$, and then the emphasis will be placed on calculations for finding the solution to the inverse problem of equation \eqref{eq:inverseProblem}.

\subsection{The Scalar Bergomi Model}
\label{sec:1factorBergomiPart1}
For $d=1$ the Bergomi market model has the volatility function,
\[\nu(t,T)  = \gamma \sigma e^{-\kappa(T-t)}\ ,\]
where $\gamma $ is a scalar constant, and $\sigma>0$ and $\kappa>0$. Define the factor process to be the Ornstein-Uhlenbeck (OU) process $X_t$ given by
\[dX_t = -\kappa X_tdt+\sigma dW_t \ ,\]
which has invariant density  
$$\omega(x)= \sqrt{\frac{\kappa}{\sigma^2 \pi} }e^{-\frac{\kappa x^2}{\sigma^2}}\ ;$$
for this model the drift and diffusion are $\mu(x)=-\kappa x$, $\sigma(x)= \sigma$. Given $X_t$, the market model's futures price is
\begin{align*}
F_{t,T} &=F^\infty \exp\left(-\frac{\gamma^2\sigma^2}{4\kappa}e^{-2\kappa(T-t)}+\gamma  e^{-\kappa(T-t)}X_t\right)\ ,
\end{align*}
where $F^\infty = \lim_{T\rightarrow \infty}F_{t,T} $ and is also a model parameter; it is straightforward to check that this expression for $F_{t,T}$ satisfies the market-model equation $dF_{t,T} =F_{t,T}\nu(t,T)dW_t$. For this model, the roll-yields of equation \eqref{eq:CMF_TH} are
\[f_\theta(X_t) = \frac{\partial}{\partial T}\log(F_{t,T})\Big|_{T=t+\theta} = \frac{\gamma^2\sigma^2}{2}e^{-2\kappa \theta} -\gamma \kappa X_t e^{-\kappa\theta}\ ,\]
and the volatilities take the form $\nu_\theta(X_t) = \gamma \sigma e^{-\kappa \theta}$. The consistency equation \eqref{eq:noArb_diffMulti} can be solved to obtain $h_\theta(x) = h_\theta(0)\exp\left(\gamma e^{-\kappa\theta} x\right)$, and it is easily verified that the solvability condition \eqref{eq:solvability} holds.

The inverse problem in \eqref{eq:inverseProblem} is solved by an eigenfunction expansion. The OU process has a complete orthogonal basis of eigenfunctions given by the Hermite polynomials. Hence, the inverse problem is solved with an eigenseries expansion like that of Section \ref{sec:eigenMethod}. 

Consider the process
\[dZ_t = -\kappa Z_tdt +\sqrt{2\kappa}dW_t \ ,\]
where $dW_tdW_t = dt$. The generator of this process is 
\[\mathcal L = \kappa\frac{\partial^2}{\partial z^2} - \kappa z\frac{\partial}{\partial z}\ ,\]
and the eigenfunctions of $\mathcal L$ satisfy equations,
\[\mathcal L\mathcal \psi_n(x)=-\kappa n \psi_n(x)\qquad\hbox{for }n=0,1,2,3,\dots\ ,\]
where each $\psi_n$ is a Hermite polynomial,
\[\psi_n(z) =(-1)^n\exp\Big(\frac{z^2}{2}\Big)\frac{d^n}{dz^n}\exp\Big(-\frac{z^2}{2}\Big)\ , \]
i.e.,
\begin{eqnarray*}
\psi_0(z)& =&1\\
\psi_1(z)&= &z\\
\psi_2(z)&= &z^2-1\\
&\vdots&
\end{eqnarray*}
Theses polynomials are orthogonal with respect to $Z_t$'s invariant measure,
\[\int_{-\infty}^\infty \psi_n(z)\psi_m(z)\omega(z)dz = n!\delta(n-m)\ ,\]
where 
\[\omega(z) =\frac{1}{\sqrt{2\pi}} \exp\Big(-\frac{z^2}{2}\Big) \ .\]
These eigenfunctions form a complete orthogonal basis in $L^2(\mathbb R;\omega)$, and are convenient because,
\[\E \left[\psi_n(Z_t)\Big|Z_0=z\right] = e^{-\kappa n t}\psi_n(z)\ .\]
The transition density for the $Z_t$'s is the following kernel,
\[\Phi_z(y,z) = \frac{1}{\tau}\int_0^\tau\frac{\partial}{\partial y}\mathbb P(Z_t\leq y|Z_0 = z)dt\ ,\]
and when applied to the Hermite polynomials,
\begin{align*}
\int_{-\infty}^\infty \Phi_z(y,z)\psi_0(y)dy &= \psi_0(z) = 1\\
\int_{-\infty}^\infty \Phi_z(y,z)\psi_n(y)dy &= \frac{1}{\tau}\int_0^\tau e^{-\kappa nt}\psi_n(z)dt = \frac{1-e^{-\kappa n \tau}}{\kappa n\tau}\psi_n(z)\qquad\forall n\geq 1\ ,
\end{align*}
which yields the eigenvalues
\begin{align*}
\lambda_0&=1\\
\lambda_n &=  \frac{1-e^{-\kappa  \tau n}}{\kappa \tau n}  \qquad\forall n\geq 1\ .
\end{align*}

\begin{figure} 
   \centering
   \includegraphics[width=5in]{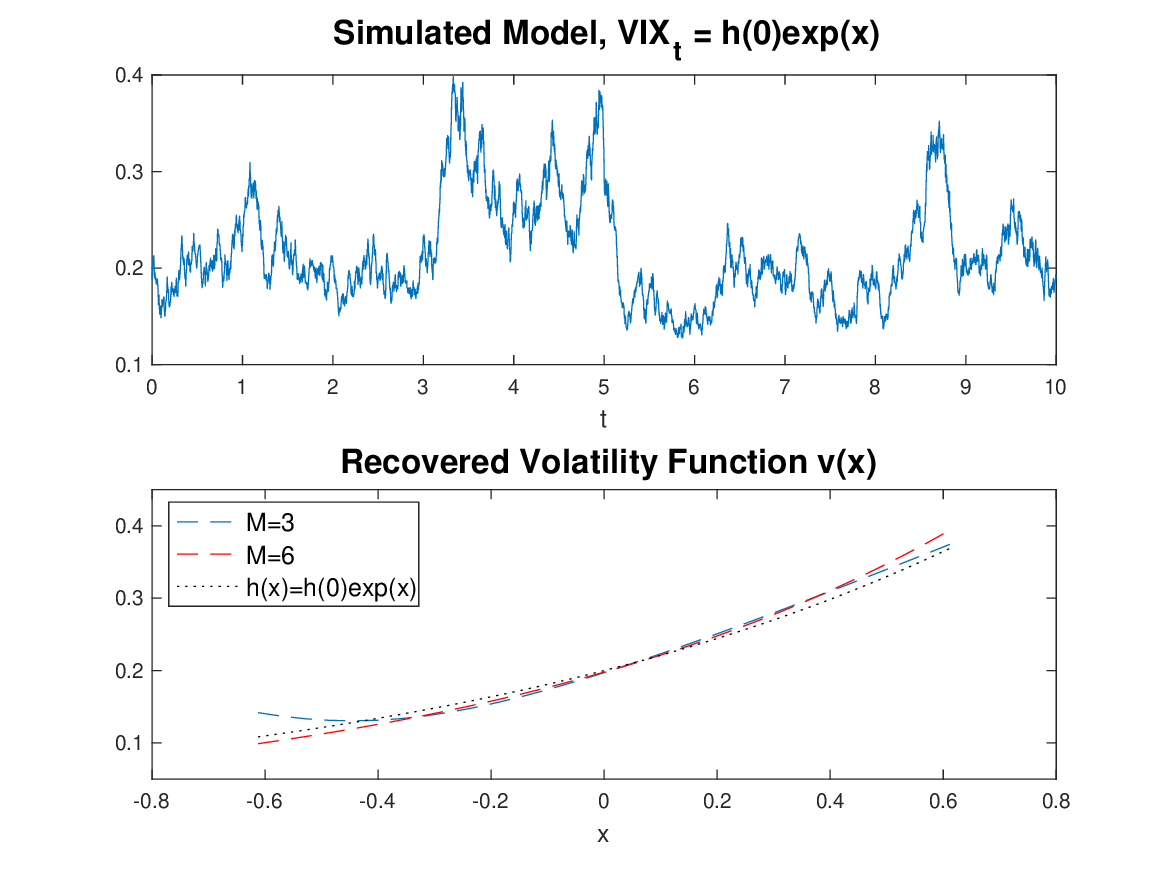} 
   \caption{For the scalar Bergomi model, the VIX is $h(x) = h(0)e^{\gamma x}$ where $h(0) =.2$, $\gamma = 1$, $\kappa=3$ and $\sigma = .5$. The mean VIX is 20.1\% and the model is 20\%.}
   \label{fig:scalarOU}
\end{figure}

For the scalar Bergomi model driven by the OU process $X_t$ with mean-reversion rate $\kappa$ and diffusion parameter $\sigma$, there is the following weak equivalence with $Z_t$,
\[X_t =_d \frac{\sigma}{\sqrt{2\kappa}}Z_t\ .\]
Define the scaled domain variance function,
\[\tilde v^2(z) =v^2\left(\frac{\sigma}{\sqrt{2\kappa} }z\right) \qquad\forall z\in\mathbb R \ ,\]
and then notice
\begin{align*}
\frac{1}{\tau}\int_0^\tau\E\left[v^2(X_t)\Big|X_0 = x\right]dt 
& = \frac{1}{\tau}\int_0^\tau \E\left[\tilde v^2(Z_t)\Big|Z_0 =  \frac{\sqrt{2\kappa }}{\sigma}x\right]dt\ .
\end{align*}
If the SVM and market model are consistent, then $\VIX_t^2=h^2(X_t)$ is given explicitly by the market model,
\[h^2(x) = h^2(0)\exp(2\gamma  x) = h^2(0)\exp\left(\frac{\sqrt{2}\gamma  \sigma}{\sqrt{\kappa}} z\right)\ .\]
Then, in terms of $z$ and the scaled eigenfunction $\tilde v^2(z)$, the solution to the inverse problem has the expansion,
\[
\tilde v^2\left(z\right)=  \sum_{n=0}^\infty a_n  \psi_n\left(z\right)\ ,
\]
and the inverse problem \eqref{eq:inverseProblem} can be written in terms of the scaled variable and variance function,
\[h^2(0)\exp\left(\frac{\sqrt{2}\gamma  \sigma}{\sqrt{\kappa}} z\right)= \frac{1}{\tau}\int_0^\tau \E\left[\tilde v^2(Z_t)\Big|Z_0 = z\right]dt=  \sum_{n=0}^\infty a_n\lambda_n  \psi_n\left(z\right)\ ,\]
for all $z\in\mathbb R$. Then using orthogonality the coefficients are,
\begin{align*}
a_n &= (-1)^n\frac{h^2(0)}{ \lambda_nn!\sqrt{2\pi}}\int_{-\infty}^\infty \exp\left(\frac{\sqrt{2}\gamma  \sigma}{\sqrt{\kappa}}z\right)\frac{d^n}{dz^n}\exp\Big(-\frac{z^2}{2}\Big)dx\\
&=\frac{h^2(0)}{\lambda_nn!}\left(\frac{\sqrt{2}\gamma  \sigma}{\sqrt{\kappa}}\right)^n\exp\Big(\frac{\gamma^2 \sigma^2}{\kappa}\Big)\ .
\end{align*}
This is clearly an expansion convergent in $L^2(\mathbb R;\omega)$ and uniformly on compact sets. Finally, in terms of $x$ the solution is,
\[ v^2\left(x\right) =\tilde v^2\left(\frac{\sqrt{2\kappa}}{\sigma}x\right) = \sum_{n=0}^\infty a_n  \psi_n\left(\frac{\sqrt{2\kappa} }{\sigma}x\right) \ .\]
This expansion is also convergent in $L^2$ and uniformly on compact sets. Numerical calculations indicate that the solution $v^2(x)$ is positive and therefore there is an acceptable volatility function. It is interesting to note that the market model for the VIX is an exponential function, leading to an exponential OU VIX futures process. However, the consistent SVM in this case does not have an exponential OU volatility function. Numerical calculations show that the instantaneous variance $v^2(x)$ has exponential-like behavior but is not an exact exponential. Figure \ref{fig:scalarOU} shows a numerical example of the simulated VIX and the recovered volatility funciton in this scalar OU example.

\subsection{The Multi-Factor Bergomi Model}
\label{sec:multiFactorBergomi}
VIX futures from the multidimensional Bergomi model are given by
\begin{align*}
\frac{dF_{t,T}}{F_{t,T}} &=   \gamma^* e^{-\mathbf k (T-t)}\sigma dW_t \ ,
\end{align*}
where $\mathbf k$ is a $d\times d$ matrix with positive eigenvalues, $\sigma$ is a $d\times d$ constant matrix, $W_t$ is $d$-dimensional uncorrelated Brownian motion, and $\gamma$ is a $d\times 1$ vector. Let $X_t$ be the multidimensional OU process given by
$$  dX_t = -\mathbf kX_tdt+ \sigma dW_t\ .$$
To ensure stationarity of $X_t$, it is enough to assume that the eigenvalues of $\mathbf k$ have positive real parts and that $(-\mathbf k,\sigma)$ is a controllable pair, i.e., 
\[\int_0^t e^{-\mathbf k u}\sigma\sigma^* e^{-\mathbf k^*u}du\qquad\hbox{is invertible for all $t>0$}\ .\]
Under these assumptions the distribution of the OU process $X_t$ will converge to a stationary state. The invariant density of $X_t$ is a $d$-dimensional Gaussian density with mean zero and $d\times d$ covariance matrix $\Sigma$ is
\begin{equation}
\label{eq:SIGMAintegral}
\Sigma = \int_0^\infty e^{-\mathbf ku}\sigma\sigma^* e^{-\mathbf k^*u}du\ ,
\end{equation}
which is finite and non-singular if the pair is controllable. From the integral formula of \eqref{eq:SIGMAintegral} it is seen that $\Sigma$ satisfies the stationary Lyapunov equation,
\begin{equation}
\label{eq:lyap}
\mathbf k \Sigma + \Sigma \mathbf k^* = \sigma \sigma^*\ .
\end{equation}
Thus, the solution to the market model's SDE for $F_{t,T}$ is
\begin{align*}
F_{t,T} &= F^\infty \exp \Big( - \frac12\gamma^* e^{-\mathbf k(T-t)} \Sigma e^{-\mathbf k^*(T-t)} \gamma  +  \gamma^* e^{-\mathbf k(T-t)}X_t  \Bigg) \ ,
\end{align*}
where $F^\infty = \lim_{T \rightarrow \infty}F_{t,T}$. For this multidimensional model, the log-future's derivative with respect to $T$ is
\[f_\theta(X_t) =\frac{\partial}{\partial T}\log(F_{t,T})\Big|_{T=t+\theta}=
\frac12\gamma^* \mathbf k e^{-\mathbf k\theta}\Sigma e^{-\mathbf k^*\theta}\gamma 
+ \frac12\gamma^* e^{-\mathbf k\theta}\Sigma e^{-\mathbf k^*\theta} \mathbf k^* \gamma
-  \gamma^*\mathbf k e^{-\mathbf k\theta}X_t \ . \]
The volatility function of equation \eqref{eq:CMF_TH} is
\[
\nu_\theta(X_t) = \gamma^* e^{-\mathbf k \theta} \sigma.
\]
The formula for the VIX is explicit and obtained from \eqref{eq:noArb_diffMulti} (up to the initial value),
\[h_\theta(x) = h_\theta (0)\exp(\gamma^*e^{-\mathbf k\theta} x)\ .\]
As in the scalar case of Section \ref{sec:1factorBergomiPart1}, it is easily verified that solvability condition \eqref{eq:solvability} holds here.

When $\mathbf k$ is diagonalizable with linearly independent eigenvectors then
the generator  $\mathcal L$ has a discrete set of eigenvalues and a complete bi-orthogonal (in general) basis of eigenfunctions given by multivariate Hermite polynomials (see \cite{ISMAIL2017,liberzonBrockett2000,withers2000}), and therefore the method of Section \ref{sec:eigenMethod} applies even though the
generator is in general not symmetric in the sense of Assumption \ref{assumption:spectralGap}.

\begin{figure}[h!] 
   \centering
   \includegraphics[width=5in]{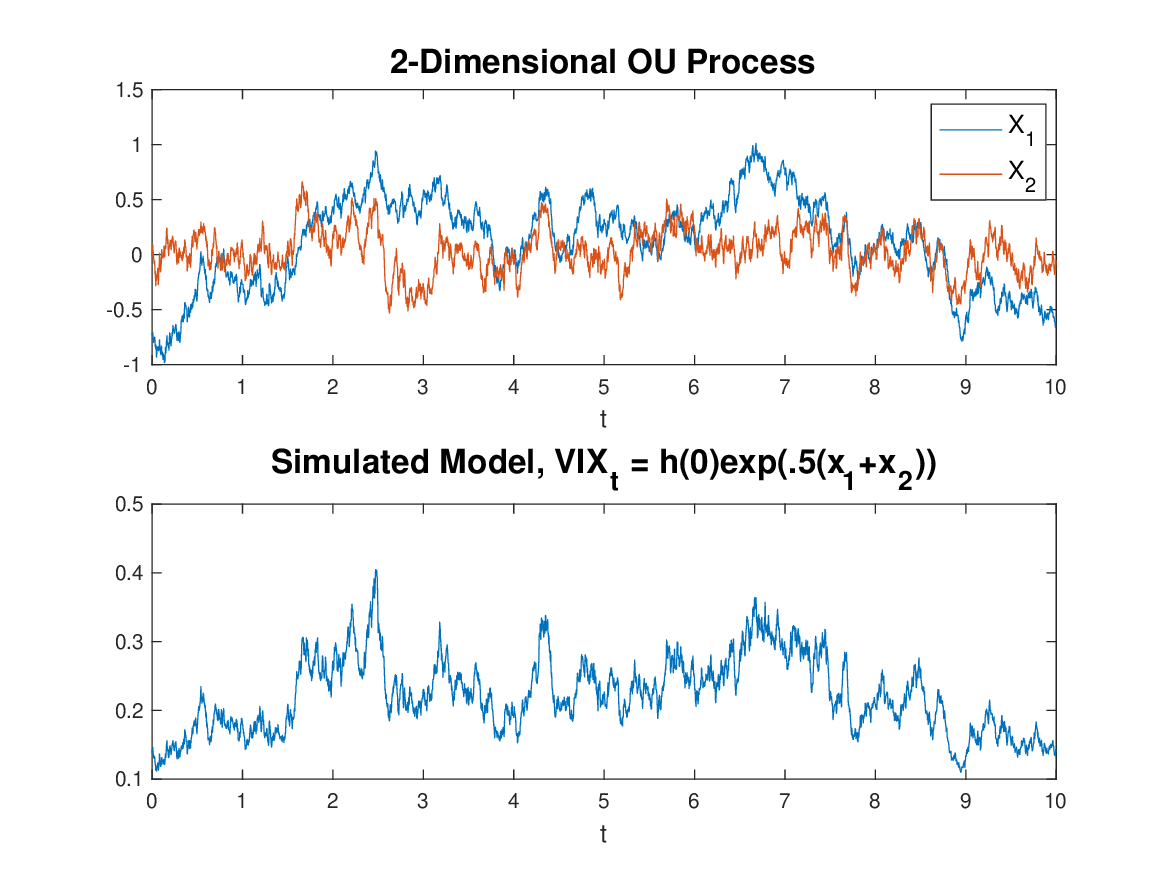} 
   \caption{For a 2-factor Bergomi model, a simulated 2-dimensional OU process $X_t=(X_t^1,X_t^2)$ and $\VIX_t  = h(0)\exp\left(\frac12(X_t^1+X_t^2)\right)$, with parameters $h(0) = .2$, $\kappa_1 =1$, $\kappa_2 = 10$, $\sigma_1 = .6$, $\sigma_2 = .8$, and $\rho = .4$. The mean VIX for this realization is $22.2\%$ and the mode is $20.0\%$. The process $X_t^1$ is \textit{persistent} because it has slower mean reversion.}
   \label{fig:twoDimOUsim}
\end{figure}

\begin{figure} 
   \centering
   \includegraphics[width=5in]{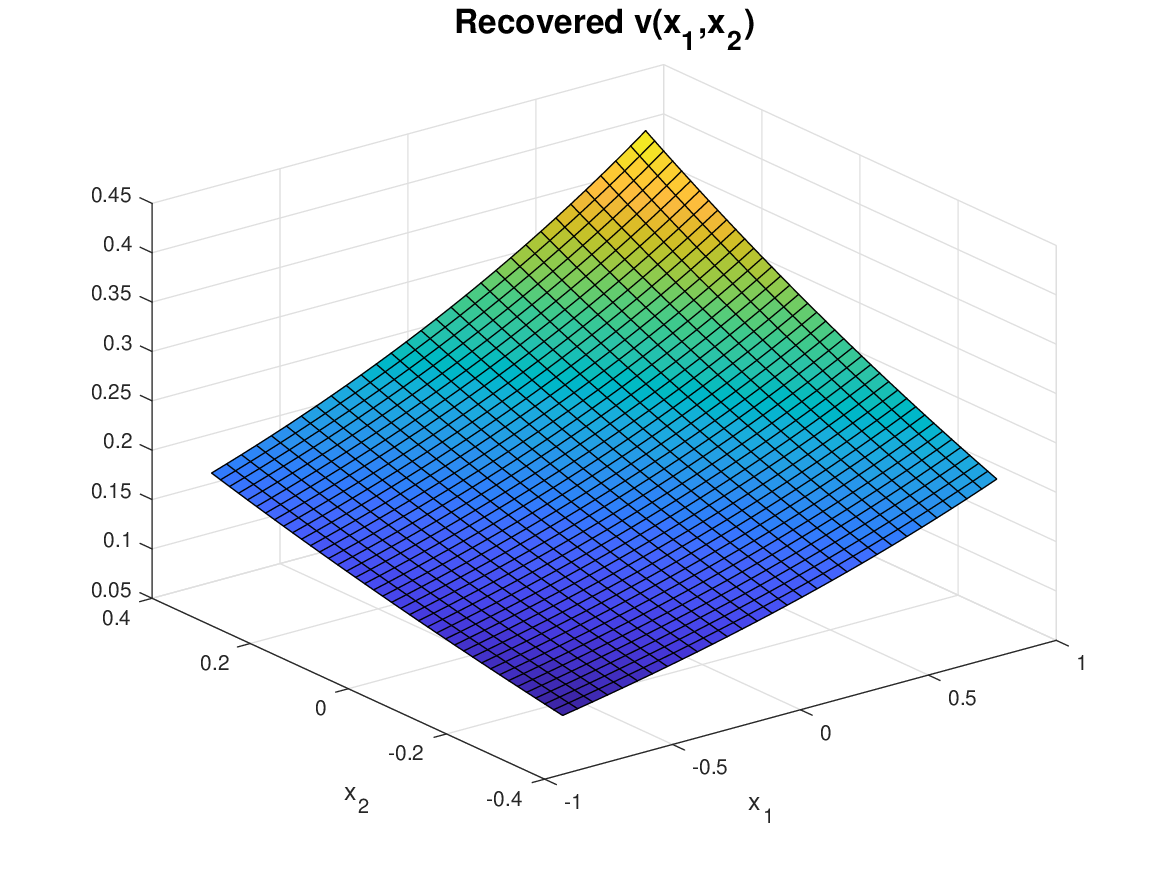} 
   \caption{An approximation of the recovered $v(x)$ from a 2-factor Bergomi model, with parameters $h(0) = .2$, $\kappa_1 =1$, $\kappa_2 = 10$, $\sigma_1 = .6$, $\sigma_2 = .8$, and $\rho = .4$. The approximated $v(x)$ uses the multivariate Hermite polynomials up to and including powers of 6. }
   \label{fig:recovered2DimOU}
\end{figure}

\begin{figure} 
   \centering
   \includegraphics[width=5in]{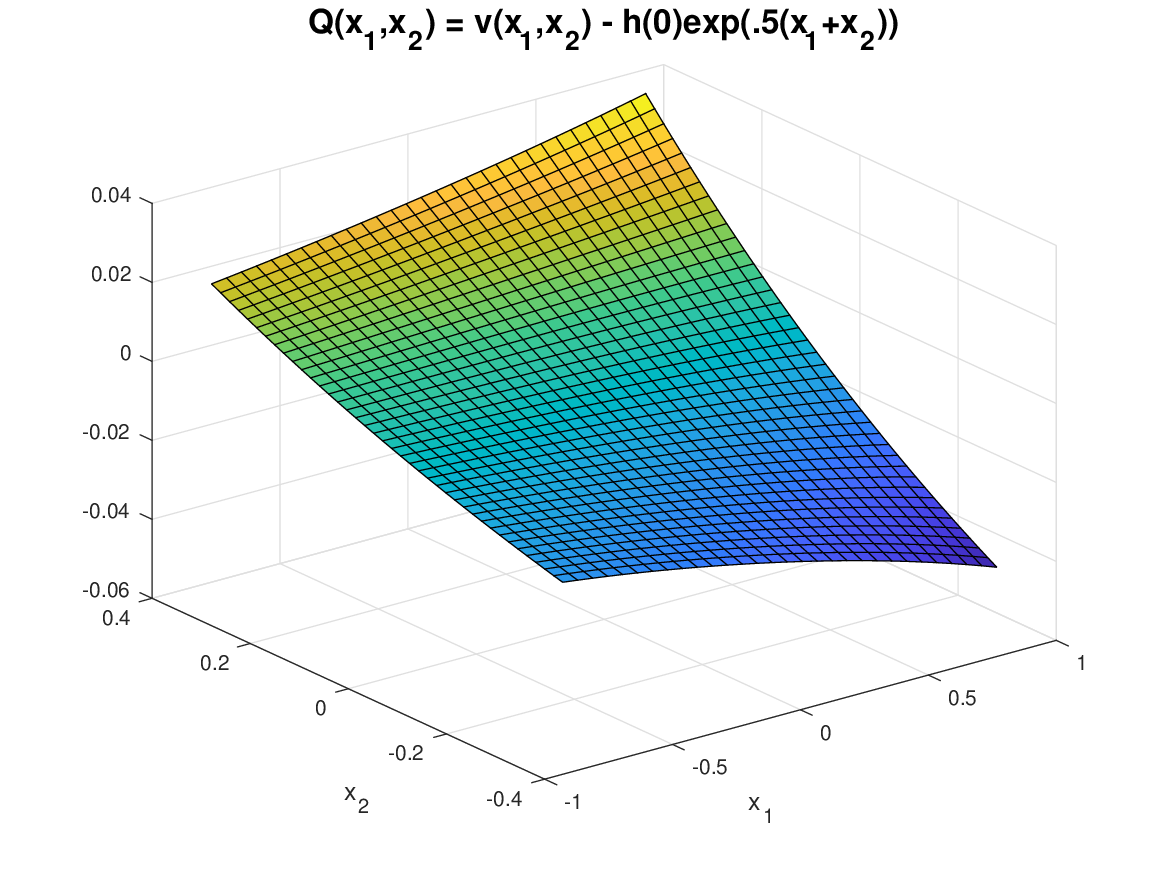} 
   \caption{For the 2-factor Bergomi model, $Q(x)$ denotes the difference between the stochastic volatility function and VIX, $Q(x) =v(x) - h(x)$ where $h(x)=h(0)\exp\left(\frac12(x_1+x_2)\right)$. The model parameters are $h(0) = .2$, $\kappa_1 =1$, $\kappa_2 = 10$, $\sigma_1 = .6$, $\sigma_2 = .8$, and $\rho = .4$. }
   \label{fig:diffOUvolsFuncs}
\end{figure}

As an example, consider the 2-dimensional model from \cite{avellanedaPapanicolaou2017,bergomi2005smile}, where the factors are

\[dX_t^i = -\kappa_i X_t^idt +\sigma_i dW_t^i\qquad\hbox{for $i=1$ and $2$,}\]
with $\kappa_i>0$ for $i=1$ and $2$, $dW_t^1dW_t^2 = \rho dt$, the VIX being
\[h(x) = h(0)\exp\left(\frac{x_1+x_2}{2}\right)\ ,\]
where it is assumed for simplicity that $\gamma = \frac{1}{2} \mathbf 1$ with $\mathbf 1 =(1,1)^*$.
The generator of $X_t$ is
\[\mathcal L=\frac12\hbox{trace}\left[\begin{pmatrix}
\sigma_1^2&\rho\sigma_1\sigma_2\\
\rho\sigma_1\sigma_2&\sigma_2^2
\end{pmatrix}\nabla\nabla^*\right] -x^* \begin{pmatrix}
\kappa_1&0\\
0&\kappa_2
\end{pmatrix}\nabla \ , \]
and the invariant density is
\[\omega(x) =\frac{1}{2\pi\sqrt{|\Sigma|}}\exp\Big(-\frac{1}{2}x^*\Sigma^{-1}x\Big) \ ,\]
where
\[\Sigma =
\begin{pmatrix}
\frac{\sigma_1^2}{2\kappa_1}&\frac{\rho\sigma_1\sigma_2}{\kappa_1+\kappa_2}\\
\frac{\rho\sigma_1\sigma_2}{\kappa_1+\kappa_2}&\frac{\sigma_2^2}{2\kappa_2}
\end{pmatrix}\ ,\]
so that \eqref{eq:lyap} holds.
Following \cite{withers2000}, the eigenfunctions $\phi_n$ for the adjoint operator $\mathcal L^*$ are
\[\phi_n(x) = \left(-\frac{\partial}{\partial x_1}\right)^{n_1}\left(-\frac{\partial}{\partial x_2}\right)^{n_2}\omega(x)\ ,\]
where $n_1$ and $n_2$ are non-negative integers; notice that $\mathcal L^*\omega = 0$. These $\phi_n$'s are the solutions to the equations
\[\mathcal L^*\phi_n = -\alpha_n\phi_n\ ,\]
where $\alpha_n = n_1\kappa_1+n_2\kappa_2 $. Then, the eigenfunctions $\psi_n$ for the operator $\mathcal L$ are multivariate Hermite polynomials, which are
\[\psi_n(x) = \frac{1}{\omega(x)}\phi_n(x)\ ,\]
and satisfy the equation
\[\mathcal L\psi_n = -\alpha_n\psi_n\ ;\]
each of these $\psi_n$'s is a polynomial of degree equal to $n_1+n_2$. In this case the transition-density kernel is
\[\Phi(y,x) = \frac{1}{\tau}\int_0^\tau\frac{\partial^2}{\partial y_1\partial y_2}\mathbb P(X_t\leq y|X_0 = x)dt\ ,\]
where $X_t\leq y$ denotes element-wise inequality, and when applied to the multivariate Hermite polynomials, similar to the scalar OU example of Section \ref{sec:1factorBergomiPart1}, there are eigenvalues
\begin{align*}
\lambda_0&=1\\
\lambda_n &=  \frac{1-e^{-\alpha_n \tau}}{\alpha_n\tau}\qquad\forall n\geq 1\ .
\end{align*}
The set of $\psi_n$'s forms a complete basis in $L^2(\mathbb R^2;\omega)$, which satisfy a bi-orthogonality relation relative to a second basis. Define this second set of basis functions to be
\[\widetilde \psi_n(x) = \frac{1}{\omega(x)}\left(-\frac{\partial}{\partial z_1}\right)^{n_1}\left(-\frac{\partial}{\partial z_2}\right)^{n_2}\omega(\Sigma z)\Bigg|_{z=\Sigma^{-1}x}\ .\]
which are bi-orthogonal in the sense that
\[\int_{\mathbb R^2} \widetilde\psi_n(x)\psi_m(x)\omega(x)dx = n_1!n_2!\delta(n-m)\ .\]
Denoting $\mathbf 1=(1,1)^*$, the inverse problem is
\[h^2(0)\exp(x^*\mathbf 1)= \frac{1}{\tau}\int_0^\tau \E\left[v^2(X_t)\Big|Z_0 = z\right]dt=\sum_{n=0}^\infty a_n\lambda_n\psi_n(x)\ ,\]
for all $x\in\mathbb R^2 $, which via the bi-orthogonality relation has the solution
\begin{align*}
a_n &= \frac{h^2(0)}{ \lambda_nn_1!n_2!}\int_{\mathbb R^2}  \exp(x^*\mathbf 1)\widetilde \psi_n(x)\omega(x)dx\\
&= \frac{h^2(0)}{ \lambda_nn_1!n_2!}\int_{\mathbb R^2}  \left[\exp(z^*\Sigma\mathbf 1)\left(-\frac{\partial}{\partial z_1}\right)^{n_1}\left(-\frac{\partial}{\partial z_2}\right)^{n_2}\omega(\Sigma z)\right]_{z=\Sigma^{-1}x}dx\\
&= \frac{h^2(0)(\Sigma_{11}+\Sigma_{21})^{n_1}(\Sigma_{22}+\Sigma_{21})^{n_2}}{ \lambda_nn_1!n_2!}\int \exp(x^*\mathbf 1)\omega(x)dx\\
&= \frac{h^2(0)(\Sigma_{11}+\Sigma_{21})^{n_1}(\Sigma_{22}+\Sigma_{21})^{n_2}}{ \lambda_nn_1!n_2!}\exp\left( \frac12\mathbf 1^*\Sigma\mathbf 1\right)\ .
\end{align*}
As with the scalar Bergomi, it is not needed to check for solvability, existence or uniqueness because the solution has eigencoefficients that are explicit. Figures \ref{fig:twoDimOUsim} and \ref{fig:recovered2DimOU} show the simulation of this 2-factor Bergomi model along with the recovered $v(x)$, which is appears to be positive, and Figure \ref{fig:diffOUvolsFuncs} looks at the difference $Q(x) = v(x)-h(x)$ to gain a sense of the differing factor sensitivities in $v(x)$ and VIX function $h(x)$. The approximated $v(x)$ uses all multivariate Hermite polynomials up to and including powers of 6, $v(x)\approx \sqrt{\sum_{\mathcal N_6}a_n\psi_n(x)}$ where $\mathcal N_6=\{n:n_1+n_2\leq 6\}$. Using only 6-degree polynomials is sufficiently accurate, as the average error in approximating is of order $10^{-6}$, i.e., $\sqrt{\frac{1}{|\mbox{x}|}\sum_{i,j}\left(\sqrt{\sum_{\mathcal N_6}a_n\psi_n(\mbox{x}_{ij})} - h(\mbox{x}_{ij})\right)^2}=\mathcal O(10^{-6})$ where $\mbox{x}_{ij}$ denotes a discrete evaluation point in $\mathbb R^2$ and $|\mbox{x}|$ denotes the total number of discrete points evaluated. Notice that $\int v^2(x)\omega(x)dx =\int h^2(x)\omega(x)dx$ (to see why multiply both sides of \eqref{eq:inverseProblem} by $\omega(x)$ and integrate). From this surface plot it can be seen that rises in the persistent factor $x_1$ have more effect on VIX than on $v(x)$ when the fast-mean-reverting factor is low (i.e., when $x_2<0$); this is seen in the corner of the surface plot where $Q(x_1,x_2)$ is most negative. This is an interesting caveat of the solution to the inverse problem, as it says that the VIX can be more persistent than instantaneous volatility, but this should not be too much of a surprise because VIX is the square-root of the expectation of a moving average of square instantaneous volatility.

\subsection{The $3/2$ Model}
\label{sec:3/2model}
Consider a market model constructed upon the squared VIX being a 3/2 process,
\[\VIX_t^2=V_t=\frac{1}{X_t}\ ,\]
where $X_t$ is a CIR process,
\[dX_t =\kappa(\bar x-X_t)dt+\sigma\sqrt{X_t}dW_t\ , \]
with $\frac{2\kappa\bar x}{\sigma^2}>2$.\footnote{The model proposed in this section is similar to that used in \cite{goardMazur2013}, wherein $\VIX_t = 1/X_t$, which could be done here as well but will require $\frac{2\bar x\kappa}{\sigma^2}>4$ to have $L^2$ integrability of the series expansion.} Applying It\^o's lemma yields
\begin{align*}
dV_t&=\frac{1}{X_t}\Big( \kappa-(\kappa\bar x-\sigma^2)\frac{1}{X_t}\Big)dt - \sigma \Big(\frac{1}{X_t}\Big)^{3/2}dW_t \\
&=V_t\Big( \kappa-(\kappa\bar x-\sigma^2)V_t\Big)dt - \sigma V_t^{3/2}dW_t \ ,
\end{align*}
from which the 3/2 power in the diffusion is seen, thus giving the process $V_t$ its name. Note that this 3/2 model is based on Assumption \ref{assumption:strongSolutions} because $X_t$'s SDE has non-Lipschitz coefficients. However, $X_t$ does have strong solutions, and the futures are $F_{t,T} = \E[\sqrt{V_T}|\mathcal F_t] $ for all $t\leq T$, which are martingales by construction. Therefore, Assumption \ref{assumption:strongSolutions} is not needed and Definition \ref{def:consistency} for consistency applies to this model. 

Consider first the normalized CIR process, which has a complete orthogonal basis of eigenfunctions for its generator, given by the generalized Laguerre polynomials. Hence, the inverse problem is again solved with an eigenseries expansion and the method of Section \ref{sec:eigenMethod} applies. Consider the normalized CIR process,
\[dZ_t = (1+\alpha - Z_t)dt+\sqrt{2Z_t}dW_t \ ,\]
where $\alpha> 0$. The generator of this process is
\[\mathcal L = z\frac{\partial ^2}{\partial z^2}+(1+\alpha-z)\frac{\partial}{\partial z}\ ,\]
and the eigenfunctions of $\mathcal L$ satisfy equations
\[\mathcal L\psi_n = -n\psi_n\qquad\hbox{for }n=0,1,2,\dots,\]
where each $\psi_n$ is a generalized Laguerre polynomial,
\begin{eqnarray*}
\psi_n(z)&=&\frac{1}{n!}e^zz^{-\alpha}\frac{d^n}{dz^n}\left(e^{-z}z^{n+\alpha}\right)\ ,
\end{eqnarray*}
that is,
\begin{eqnarray*}
\psi_0(z)& =&1\\
\psi_1(z)&=& -z+\alpha+1\\
\psi_2(z)&=& \frac12z^2-(\alpha+2)z+\frac12(\alpha+2)(\alpha+1)\\
&\vdots &
\end{eqnarray*}
These polynomials are orthogonal with respect to $Z$'s invariant measure,
\[\int_0^\infty \psi_n(z)\psi_m(z)\omega(z)dz  =c_n\delta(n-m)\ ,\]
where 
$$c_n = \frac{\Gamma(n+\alpha+1)}{n!~\Gamma(\alpha+1)}\ ,$$ 
and
\[\omega(z) = \frac{1}{\Gamma(\alpha+1)}z^{\alpha}\exp(-z)\ ,\]
with $\Gamma(\alpha)$ the Gamma function evaluated at $\alpha>1$. These eigenfunctions form a complete orthogonal basis in $L^2(\mathbb R^+;\omega)$, and are convenient because
\[\E \left[\psi_n(Z_t)\Big|Z_0=z\right] = e^{-n t}\psi_n(z)\ .\]

\begin{figure} 
   \centering
   \includegraphics[width=5in]{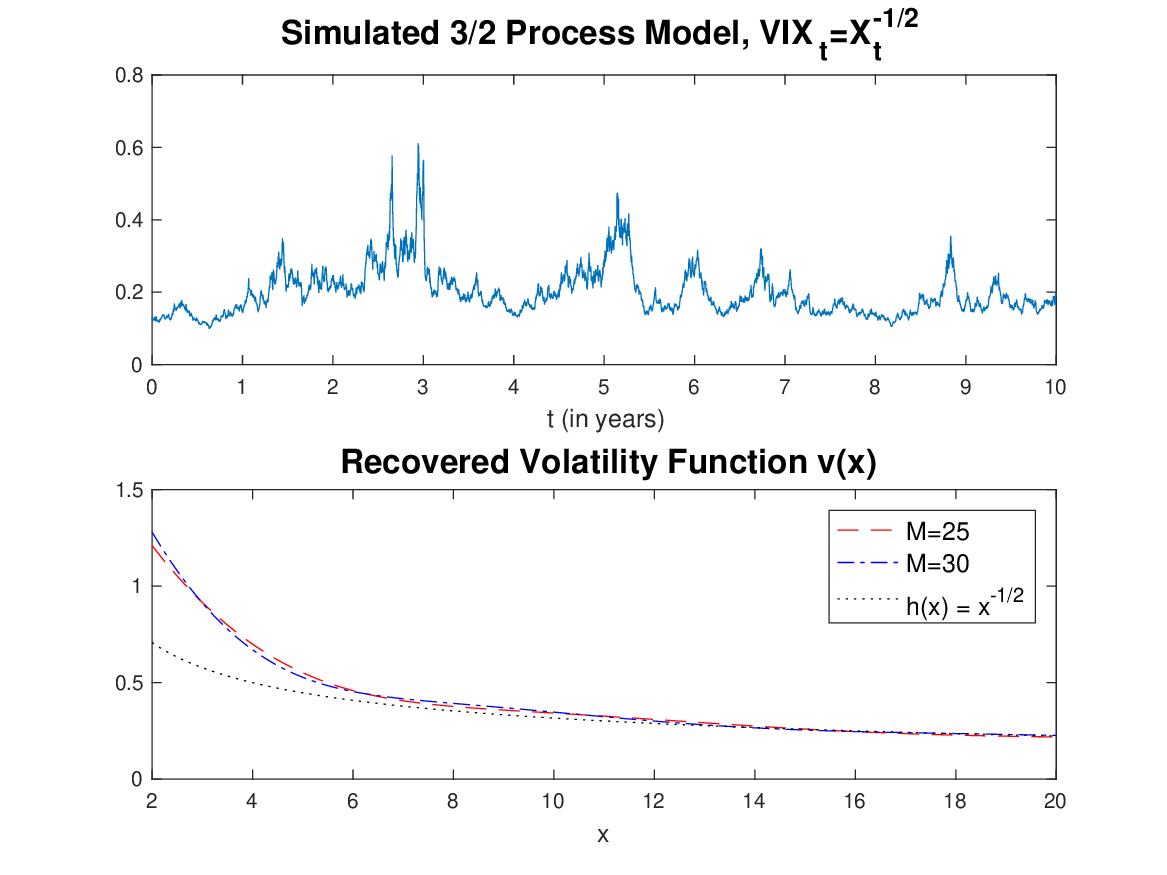} 
   \caption{The simulation of the 3/2 process and the recovered volatility function for the market model having $\VIX_t^2=\frac{1}{X_t}$, where $X_t$ is a CIR process. In the legend of the bottom plot, $M=25$ indicates an approximation using twenty-five Laguerre polynomials, $M=30$ using thirty polynomials, and $h(x) = 1/\sqrt{x}$ is the market model's VIX function. In both plots the CIR parameters are $\kappa = 4$, $\bar x = 30$ and $\sigma = 6.9282$; this yields $\alpha = 4$. For these parameters the CIR process is between 10 and 50 approximately 95\% of the time.}
   \label{fig:threeHalfs}
\end{figure}

For the CIR process $X_t$ defined above, there is the following weak equivalence with a scaled $Z_t$,
\[X_t = _d \frac{\sigma^2}{2\kappa }Z_{\kappa t}\ ,\]
with $\alpha = \frac{2\bar x\kappa}{\sigma^2}-1$. Define also the scaled domain variance or volatility function,
\[\tilde v^2(z) =v^2\left(\frac{\sigma^2}{2\kappa }z\right) \qquad\forall z>0 \ ,\]
and then notice
\begin{align*}
\frac{1}{\tau}\int_0^\tau\E[v^2(X_t)|X_0 = x]dt
& = \frac{1}{\tau}\int_0^\tau \E\left[\tilde v^2(Z_{\kappa t})\Big|Z_0 = \frac{2\kappa }{\sigma^2}x\right]dt\ .
\end{align*}
Therefore it is useful to define the kernel for the $Z_t$'s, $\Phi_z(y,z) = \frac{1}{\tau}\int_0^\tau\frac{\partial}{\partial y}\mathbb P(Z_{\kappa t}\leq y|Z_0 = z)dt$, and when applied to the Laguerre polynomials, similar to the scalar OU example, 
\begin{align*}
\int_0^\infty \Phi_z(y,z)\psi_0(y)dy &= 1\\
\int_0^\infty \Phi_z(y,z)\psi_n(y)dy &= \frac{1}{\tau}\int_0^\tau e^{-\kappa nt}\psi_n(z)dt = \frac{1-e^{-\kappa n \tau}}{\kappa n\tau}\psi_n(z)\qquad\forall n\geq 1\ ,
\end{align*}
there are the eigenvalues,
\begin{align*}
\lambda_0&=1\\
\lambda_n &=  \frac{1-e^{-\kappa n \tau}}{\kappa n\tau}\qquad\forall n\geq 1\ .
\end{align*}
Hence, if the SVM and market model are consistent, then $\VIX_t^2=h^2(X_t)$ is given explicitly by the market model,
\[h^2(x) = \frac{1}{x} = \frac{2\kappa}{\sigma^2z}\ ,\]
which is in $L^2(\mathbb R^+;\omega)$ if $\alpha>1$. Then, in terms of $z$ and the scaled function $\tilde v^2(z)$, the solution to the inverse problem has the expansion, 
\[ \tilde v^2\left(z\right) = \sum_{n=0}^\infty a_n \psi_n\left(z\right) \ ,\]
and therefore
\[\frac{2\kappa}{\sigma^2z}= \frac{1}{\tau}\int_0^\tau \E\left[\tilde v^2(Z_{\kappa t})\Big|Z_0 = z\right]dt=\sum_{n=0}^\infty a_n\lambda_n\psi_n(z)\ ,\]
for all $z> 0 $. Using orthogonality, the coefficients are
\begin{align*}
a_ n &=\frac{2\kappa}{\sigma^2\lambda_n c_n n! \Gamma(1+\alpha)} \int_0^\infty \frac{1 }{  z}\frac{d^n}{dz^n}\left(e^{-z}z^{n+\alpha}\right)dz\\
&=\frac{2\kappa }{\sigma^2\lambda_nc_n\Gamma(1+\alpha)} \int_0^\infty e^{-z}z^{\alpha-1}dz\\
&=\frac{2\kappa }{\sigma^2\lambda_nc_n\Gamma(1+\alpha)}\Gamma(\alpha)\\
&=\frac{2\kappa\Gamma(\alpha)n! }{\sigma^2\lambda_n\Gamma(n+\alpha+1)}\ . 
\end{align*}
For $n$ large there is the behavior $a_n \approx n^{-\alpha +1}$, which requires $\alpha > 2$ for square integrability of the expansion
of $\tilde v^2(z)$. Finally, in terms of $x$, the solution is
\[ v^2\left(x\right)= \sum_{n=0}^\infty a_n \psi_n\left(\frac{2\kappa}{\sigma^2}x\right) \ .\]
Figure \ref{fig:threeHalfs} shows a simulation of the 3/2 process and the two approximations of the recovered function $v(x)$ using 25 and 30 Laguerre polynomials. In the figure, the simulation is run for 10 years with time step $\Delta t = 1/365$, and produces empirical statistics $\frac{1}{N}\sum_{i=1}^N\VIX_{t_i} = 19.89\%$ and $\hbox{mode}_{i\leq N}(\VIX_{t_i}) = 16.0\%$ (in the summation $N=10\times 365 = 3,650$). The figure shows a recovered $v^2(x)$ from which it is clear that, compared to the VIX, instantaneous volatility is more affected by low values of $X_t$; i.e., stochastic volatility is more sensitive to the left-hand tail distribution of $X_t$. Note also from the figure that the numerical solution is positive.

\subsection{The Double Nelson Model}
\label{sec:doubleNelson}
Consider the 2-dimensional mean reverting process $X_t = (X_t^1,X_t^2)$ with dynamics,
\begin{align}
\label{eq:DNelson1}
dX_t^1&=\kappa_1(X_t^2-X_t^1)dt+\sigma_1X_t^1dW_t^1\\
dX_t^2&=\kappa_2(\bar x-X_t^2)dt+\sigma_2X_t^2dW_t^2 \nonumber\ ,
\end{align}
where $\bar x>0$, $\kappa_1>0$, $\kappa_2>0$, and $dW_t^1dW_t^2=\rho dt$. This is the double Nelson model, which is the continuous-time limit of a double GARCH model. Defining the VIX to be
\[\VIX_t =h(X_t)= X_t^1\ ,\]
the futures curve $F_{t,T}=\E[h(X_T) | X_t]$ is
\[F_{t,T} = X_t^1e^{-\kappa_1(T-t)} + \bar x(1-e^{-\kappa_1(T-t)} )+( X_t^2-\bar x)\frac{\kappa_1(e^{-\kappa_2(T-t)}-e^{-\kappa_1(T-t)})}{\kappa_1-\kappa_2}\ .\]
This is a market model for which the inverse problem will look to find $v^2(x)$ from an SVM driven by the same factors $X_t^1$ and $X_t^2$. 

This model's infinitesimal generator is not symmetric and so the general theory of Theorem \ref{thm:main} does not apply directly. However, the factor process satisfies a linear system of stochastic differential equations for which there are closed equations for moments of all orders, and so the solvability condition given by \eqref{eq:solvability} from Section \ref{sec:generalSolvability} can be applied.

The zero-maturity roll yield is
\[  \frac{\partial}{\partial T}\log(F_{t,T})\Big|_{T=t}  = f(x) = \kappa_1\left(\frac{x_2}{x_1}-1\right)\ ,\]
the volatility is $\nu(t,t)=\sigma_1$, and so the solvability condition of equation \eqref{eq:solvability} is
\begin{equation}
\label{eq:DMRsolvability}
2\kappa_1\left<x_1(x_2-x_1)\right> = -\left<\sigma_1^2x_1^2\right>\ .
\end{equation}
Invariant moments can be calculated using It\^o's lemma and then taking expectations,
\begin{align*}
\left<x_2\right>&=\bar x\\
\left<x_1\right>&=\bar x\\
\left<x_2^2\right>&=\frac{2\kappa_2\bar x^2}{2\kappa_2-\sigma_2^2}\\
\left<x_1x_2\right>&=\frac{\kappa_2\bar x^2+\kappa_1\left<x_2^2\right>}{\kappa_1+\kappa_2-\rho\sigma_1\sigma_2}\\
\left<x_1^2\right>&=\frac{2\kappa_1\left<x_1x_2\right>}{2\kappa_1-\sigma_1^2}\ . 
\end{align*}
Hence, provided that $2\kappa_1-\sigma_1^2>0$ and $2\kappa_2-\sigma_2^2>0$ to ensure that $X_t$ has finite (invariant) second moments, and that $\left<x_1x_2\right>$ is finite,
\[\kappa_1+\kappa_2-\rho\sigma_1\sigma_2 > \frac12(\sigma_1,\sigma_2)\begin{pmatrix}1&-\rho\\-\rho&1\end{pmatrix}\begin{pmatrix}\sigma_1\\\sigma_2\end{pmatrix}>0\ ,\]
it follows that equation \eqref{eq:DMRsolvability} holds. 

The inverse problem is
\[h^2(x) =  \E\left[\frac{1}{\tau}\int_0^\tau v^2(X_u)du\Big|X_0=x\right] \ ,\]
with $h^2(x) = x_1^2$, and is solved explicitly by looking for the solution in the form,
\[
v^2(x)= a_{11}x_1^2 + a_{12} x_1 x_2 + a_{22} x_2^2 + b_1 x_1 + b_2 x_2 + c\ .
\]
The coefficients $a_{ij}$ and $b_i$ for $i,j=1,2$ are obtained by solving explicitly for the moments $u_{11}(t)=\E[(X_t^1)^2| X_0=x],~u_{12}(t)=\E[X_t^1 X_t^2 | X_0=x]\ , \ldots \ ,$ 
which satisfy a linear system of ordinary differential equations obtained by Ito's formula from the stochastic
differential equations of the factor process \eqref{eq:DNelson1},
\begin{align*}
\frac{du_{11}}{dt} &= -(2\kappa_1 -\sigma_1^2) u_{11} + 2\kappa u_{12}\ ,~~~~ u_{11}(0)=x_1^2\\
\frac{du_{12}}{dt} &= -(\kappa_1 + \kappa_2 -\sigma_1 \sigma_2 \rho) u_{12} + \kappa_1 u_{22} + \kappa_2 \bar x u_1\ ,~~~~u_{12}(0)=x_1 x_2\\
\frac{du_{22}}{dt} &= -(2\kappa_2 -\sigma_2^2) u_{22} + 2\kappa \bar x u_2\ ,~~~~u_{22}(0)= x_2^2\\
\frac{du_{1}}{dt} &= \kappa_1(u_2-u_1)\ ,~~~~u_1(0)= x_1\\
\frac{du_{2}}{dt} &= \kappa_2(\bar x - u_2)\ ,~~~~u_2(0)=x_2\ .
\end{align*}
Note that the invariant moments obtained above are simply the limit of these moments as $t\to \infty$, and this
requires that the relations between $\kappa_1,\kappa_2,\sigma_1,\sigma_2,\rho$ introduced above hold here too. Hence,
\begin{align*}
x_1^2 &= \frac{1}{\tau}\int_0^\tau\E[v^2(X_t)|X_0=x]dt \\
&= \frac{1}{\tau}\int_0^\tau \big(a_{11}u_{11}(t) + a_{12} u_{12}(t) + a_{22} u_{22}(t) + b_1 u_1(t)+ b_2 u_2(t) + c\big) dt\ ,
\end{align*}
and by adjusting the coefficients $a_{11},a_{12}, \ldots$ the solution is found to be $v^2(x)=v^2(x_1,x_2)$ for $x_1\geq 0$ and $x_2\geq 0$, which is
a quadratic polynomial in $(x_1,x_2)$. However, this solution will not be non-negative and therefore it is not acceptable for an SVM.

To see how the solution $v^2(x_1,x_2)$ can go negative, consider the simplified inverse problem,
\[
x_1 = \frac{1}{\tau}\int_0^\tau\E[v^2(X_t)|X_0=x]dt\ ,
\]
which requires only a linear expression for its solution
\[
v^2(x)= b_1 x_1 + b_2 x_2 + c\ .
\]
Thus
\begin{align*}
x_1 &= \frac{1}{\tau}\int_0^\tau\E[v^2(X_t)|X_0=x]dt \\
&= \frac{1}{\tau}\int_0^\tau \big(b_1 u_1(t)+ b_2 u_2(t) + c\big) dt\ ,
\end{align*}
with
\begin{align*}
u_1(t) &=e^{-\kappa_1 t}x_1 + \frac{\kappa_1(e^{-\kappa_2 t} -e^{-\kappa_1 t})}{\kappa_1 -\kappa_2} x_2
+(1-e^{-\kappa_1 t}) \bar x - \frac{\kappa_1(e^{-\kappa_2 t} -e^{-\kappa_1 t})}{\kappa_1 -\kappa_2} \bar x\ ,\\
u_2(t) &= e^{-\kappa_2 t}x_2 +(1-e^{-\kappa_2 t}) \bar x\ .
\end{align*}
Inserting these expressions and doing the time averaging it is seen that in order to solve the inverse problem it must be that
\[
b_1 = \frac{\kappa_1 \tau}{1- e^{-\kappa_1 \tau}}\ ,
\]
so that the coefficient of $x_1$ on the right is one. Then taking $b_2$ to make the coefficient of $x_2$ equal
to zero, this leads to
\[
b_2 = - b_1  \frac{\kappa_2 \tau}{1- e^{-\kappa_2 \tau}}  \frac{1}{\tau}\int_0^\tau \frac{\kappa_1(e^{-\kappa_2 t} -e^{-\kappa_1 t})}{\kappa_1 -\kappa_2} dt\ .
\]
After solving for $b_1$ and $b_2$, the constant $c$ equals to the remaining terms. Finally it is seen that $b_2$ is negative for any $\kappa_1,\kappa_2$, and this makes the solution $v^2(x)= b_1 x_1 + b_2 x_2 + c, ~x_1 \geq 0,x_2 \geq 0,$ take
negative values for $x_1$ near $0$ and $x_2$ large. This indicates that there cannot be consistency in the sense of Definition \ref{def:consistency}.

\subsection{Non-Negative Solutions for Brownian Motion Factor}
\label{sec:brownianMotion}

Consider another example that does not have the stationarity of Assumptions \ref{assumption:invariantDensity} and \ref{assumption:spectralGap}, but instead is a market model driven by Brownian motion $Z_t$. This is an example that has a general condition on $h^2$ to ensure non-negativity of the recovered volatility function.

The inverse problem is
\[h^2(z) = \frac{1}{\tau}\int_0^\tau \mathbb Ev^2(z+Z_t)dt\ ,\]
where $Z_t$ is standard Brownian motion. The Fourier transform is used to solve this problem. The space to consider is $L^2(\mathbb R,dz)$, and the Fourier elements are
\[\psi(k,z) =\frac{1}{\sqrt{2\pi}}e^{i kz}\ ,\]
which have generalized orthogonality with the delta function $\delta(k-k') = \frac{1}{2\pi}\int e^{i  k'z}e^{-i kz}dz$. The market model's VIX function and the SVM function have Fourier transforms
\begin{align*}
	\widehat{h^2}(k) &=\frac{1}{\sqrt{2\pi}}\int e^{-i kz}h^2(z)dz\\
	\widehat{v^2}(k)&=\frac{1}{\sqrt{2\pi}}\int e^{-i kz}v^2(z)dz \ .
\end{align*}
The Fourier basis is used to transform the inverse problem,
\begin{align*}
	\widehat{h^2}(k)&=\frac{1}{\tau\sqrt{2\pi}}\int_0^\tau\mathbb E e^{i kZ_t} \int e^{-i k(z+Z_t)}v^2(z+Z_t)dz dt\\
	&=\frac{2\widehat{v^2}(k)}{k^2\tau}\left(1- e^{-\frac{\tau k^2}{2}} \right)\ .
\end{align*}
Hence, the solution to the problem is
\[\widehat{v^2}(k)= \frac{k^2\tau}{2(1- e^{-\frac{\tau k^2}{2}} )}\widehat{h^2}(k)\ .\]
If $k^2\widehat{h^2}(k)$ is in $L^2(\mathbb R,dk)$, then Parseval's identity says the solution $v^2(z)$ is in $L^2(\mathbb R,dz)$,
\[\|v^2\|^2= \int \left| \frac{k^2\tau\sqrt{2\pi}}{2(1- e^{-\frac{\tau k^2}{2}} )}\widehat{h^2}(k)\right|^2 dk<\infty\ . \]
If $\widehat{v^2}(k)$ is continuous and positive definite, that is, if for any $k_\ell\in\mathbb R$ and $c_\ell\in\mathbb C$ for $\ell=1,2,3,\dots,M$ for $M$ any positive integer, 
\[\sum_{\ell,\ell'=1}^Mc_\ell\bar c_{\ell'}\widehat{v^2}(k_\ell-k_{\ell'})\geq 0 \ ,\]
then Bochner's theorem applies \cite{reed2012methods} and $v^2(z) = \frac{1}{\sqrt{2\pi}}\int e^{i kz}\widehat{v^2}(k)dk$ is non negative. This is a general criterion for the solution to the inverse problem to be non-negative, but this application of Bochner's theorem is special to the case of Fourier eigenfunctions.

To further illustrate, consider the specific example 
\[h^2(z) = c+\gamma z^2\ ,\]
and with $Z_t$ a standard Brownian motion. For inverse problem
\[h^2(z) = \frac{1}{\tau}\int_0^\tau \mathbb E[v^2(z+Z_t)|Z_0 = 0]\ ,\]
it is easy to check,
\[c+\gamma z^2 = \frac{1}{\tau}\mathbb E\int_0^\tau \left(c+\gamma (z+Z_t)^2-\frac{\tau\gamma}{2}\right)dt\ .\]
Hence, there is non-negative solution if $c\geq \frac{\tau\gamma}{2}$.

\section{Non-Markovian Market Models}
\label{sec:nonMarkov}

Let $h_\theta(x)$ denote the CMFs derived from the Markovian SVM. Suppose that Assumption \ref{assumption:markovX} does not hold so that it's possible to have non-Markovian dynamics. Then to check for the consistency of Definition \ref{def:consistency}, there are the following pair of equations that are the generalization of \eqref{eq:noArb_driftMulti} and \eqref{eq:noArb_diffMulti},
\begin{eqnarray}
\label{eq:noArb_driftMultiGEN}
\frac{\frac{1}{2}\hbox{trace}[\sigma \sigma^*(X_t)\nabla\nabla^*]  h_\theta(X_t)+\mu^*(X_t)\nabla h_\theta(X_t)}{h_\theta(X_t)}&=&Y_t^\theta \\
\label{eq:noArb_diffMultiGEN}
\frac{\sigma^*(X_t)\nabla h_\theta(X_t)}{h_\theta(X_t)}&=&\nu_\theta^*(t) \ ,
\end{eqnarray}
where $Y_t^\theta $ is the roll yield as shown in equation \eqref{eq:CMF_dF}. From equations \eqref{eq:noArb_driftMultiGEN} and \eqref{eq:noArb_diffMultiGEN} it should be clear that a Markovian representation of the market model must be imposed. Namely, $Y_t^\theta = f_\theta(X_t)$ where $f_\theta(X_t)$ equals the left-hand side of equation \eqref{eq:noArb_driftMultiGEN}, and $\nu_\theta(t)=\nu_\theta(X_t)$ where $\nu_\theta(X_t)$ equals the transpose of the left-hand side of equation \eqref{eq:noArb_diffMultiGEN}. 

\subsection{Scalar Consistency with Constant $\nu_\theta(t)$}
Consider the case where $X_t$ and $W_t$ in equation \eqref{eq:SVM_dX} are scalar processes. Suppose that $\nu_\theta$ is a scalar, constant deterministic function,
\[\nu_\theta(t) = \nu_\theta\in\mathbb R^1\qquad\forall t\ .\]
Then solving equations \eqref{eq:noArb_driftMultiGEN} and \eqref{eq:noArb_diffMultiGEN}  leads to the following VIX futures and roll yields, 
\begin{align}
\label{eq:noArb_f}
f_\theta(x)&= \nu_\theta\frac{\mu(x)+\frac{1}{2}\sigma(x)\Big(\nu_\theta- \frac{d}{dx}\sigma (x)\Big)}{\sigma(x)}\ ,\\
\label{eq:hFormula}
h_\theta(x) &= h_\theta(x_0) \exp\left(\nu_\theta\int_{x_0}^x\frac{dy}{\sigma(y)}\right)\qquad\forall\theta\geq 0\ .
\end{align}
It is assumed in this equation that $\sigma(x)$ is strictly positive and its inverse is integrable.

\subsection{An Inconsistent Example}
There are non-trivial cases where there is a violation of the scalar consistency formula of equation \eqref{eq:noArb_f}. For example, suppose there is algebraic decay in the market model's volatility function,
\[\nu_\theta = \frac{\gamma}{1+\theta}\ .\]
Then the SDE for the CMF can be computed via It\^o's lemma, which yields the following roll yield, 
\[Y_t^\theta = \frac{\gamma^2}{(1+\theta)^2}- \int_{-\infty}^t\frac{\gamma}{(1+t+\theta-u)^2}dW_u\ .\]
There is no function of the Markov process $X_t$ that can equal this process almost surely, as $Y_t^\theta$ itself is not a Markov process. Hence, formula \eqref{eq:noArb_f} cannot hold.


\section{Summary and Conclusion}

The achievement of this paper is the derivation of a consistent SVM for the SPX given a market model for the VIX. The main result is Theorem \ref{thm:main}, which gives conditions for the unique determination of the volatility function of the SVM from a VIX function given by the market model, provided both models are driven by the same underlying stationary ergodic factor process. The theorem's conditions involve moments of the VIX function, the uniqueness of the invariant measure of the factor process, and require that the operator semi-group have a spectral gap. At the time of this article there are no known structural conditions that will make the resulting volatility function non-negative, and therefore no theoretical guarantees for consistency can be made. There are special cases where positivity can be guaranteed, such as models where $X_t$ is a Brownian motion. Detailed analysis and numerical calculations for several market models indicate that for the commonly used Bergomi market models (Sections \ref{sec:1factorBergomiPart1} and \ref{sec:multiFactorBergomi}) the volatility function appears to be positive. For another market model where square VIX is the reciprocal of a CIR process (Section \ref{sec:3/2model}), the volatility is again shown numerically to be positive. The double Nelson model in Section \ref{sec:doubleNelson} is a counter example, wherein the market model's factor process is a linear SDE that is stationary ergodic, but the inverse problem leads to a (unique) volatility function that cannot be everywhere non-negative. Positivity can be guaranteed for the example in Section \ref{sec:brownianMotion} because the factor process is Brownian motion, and therfore Bochner's theorem gives general conditions for non-negativity. 

Future problems to consider include general results for non-negativity of recovered volatility functions under the OU and the CIR processes, and also to generalize this inverse problem formulation to jump-diffusion models like that of \cite{duffie2000transform}. From a computational standpoint, it would be worth solving the inverse problem not as an exact equality, but instead as a minimization subject to the constraint that $v^2\geq 0$. Then, under a loosening of conditions for consistency, it could be possible that this constrained minimization will produce useful SVMs from a broader class of market models.


\bibliography{inconsistencyNote.bib}
\bibliographystyle{plain}

\end{document}